\documentclass[twocolumn,showpacs,fleqn,nobibnotes]{revtex4}

\usepackage{amsmath}
\usepackage{graphicx}
\usepackage{float}
\usepackage{subfigure}

\def\lsim{\raise0.3ex\hbox{$<$\kern-0.75em\raise-1.1ex\hbox{$\sim$}}}
\def\gsim{\raise0.3ex\hbox{$>$\kern-0.75em\raise-1.1ex\hbox{$\sim$}}}

\begin{document}

\title{Light vector meson photoproduction in hadron-hadron and nucleus-nucleus  collisions at the energies available at the CERN Large Hadron Collider}
\author{G. Sampaio dos Santos and M.V.T. Machado}

\affiliation{High Energy Physics Phenomenology Group, GFPAE  IF-UFRGS \\
Caixa Postal 15051, CEP 91501-970, Porto Alegre, RS, Brazil}

\begin{abstract}
In this work we analyse the theoretical uncertainties on the predictions for  the photoproduction of light vector mesons in coherent $pp$, $pA$ and $AA$  collisions at the LHC energies using the color dipole approach. In particular, we present our predictions for the rapidity distribution for $\rho^0$ and $\phi$ photoproduction and perform an analysis on the uncertainties associated to the choice of vector meson wavefunctionand the phenomenological models for the dipole cross section. Comparison is done with the recent ALICE analysis on coherent production of $\rho^0$ at 2.76 TeV in PbPb collisions. 

\end{abstract}

\maketitle

\section{Introduction} 

The exclusive photoproduction  of vector mesons has been investigated recently both experimentally and theoretically \cite{LHCb1V,LHCb2V,ALICE1V,ALICE2V,ALICE3V}. In particular, the light vector mesons as $\rho$ and $\phi$ have not a perturbative scale associated to the process in photoproduction limit and so they test the non-perturbative regime of QCD. The transition between the perturbative hard treatment and the soft regime can be addressed by the so-called saturation approaches \cite{hdqcd} within the color dipole formalism \cite{nik}. In those formalisms a saturation scale characterizes the limitation on the maximum phase-space parton density that can be reached in the hadron wavefunction. In such framework, the typical scale driven the dynamics of light meson production is the saturation scale and the photons can be considered as color dipoles in the mixed light cone representation, where their transverse size can be considered frozen during the interaction \cite{Nemchik:1996pp}. The corresponding scattering process is characterized by the color dipole cross section describing the interaction of those color dipoles with the nucleon or nucleus targets. Accordingly, the $q\bar{q}$ fluctuations, i.e. the color dipoles, of the incoming quasi-real photon interacts with the  target via the dipole cross section and the result is projected in the wavefunctionof the observed hadron. At high energies, i.e. small-$x$ region, it is expected a transition between the regime described by the linear dynamics of emissions chain and a new regime where the physical process of recombination of partons turn out to be crucial.  The transition is driven by the saturation scale $Q_{\mathrm{sat}}\propto A^{1/3}x^{\lambda}$, which is typically enhanced in the scattering on  nuclei targets \cite{hdqcd}. 

The approach shortly described above has done a good job in describing the vector meson photo and electroproduction at the DESY-HERA energy regime considering a proton target (see, e.g. Ref. \cite{Anelise}). The possibility for investigation of the  photo-nuclear production in similar energies was provided  by the RHIC measurements on $\rho$ and $J/\Psi$ production considering the coherent gold-gold heavy ion collisions \cite{RHIC,RHICpsi}. For a long time has been proposed the analysis of coherent collisions in hadronic interactions as an alternative way to investigate the QCD dynamics at high energies
\cite{vicmag_upcs1,vicmag_upcs2,vicmag_hq,vicmag_mesons_per,vicmag_prd,vicmag_pA,vicmag_difper,vicmag_quarkonium,vicmag_rho,vicmag_ane}. The basic idea in coherent  hadronic collisions is that
the  cross section for a given process can be factorized in
terms of the equivalent flux of photons into the hadron projectile and
the photon-photon or photon-target production cross section \cite{upcs}.
The main advantage of using colliding hadrons and nuclear beams for
studying photon induced interactions is the high equivalent photon
energies and luminosities achieved at RHIC and LHC.  Consequently, studies of $\gamma p$  or $\gamma A$ interactions
at the LHC could provide valuable information on the QCD dynamics at high energies. The main point here is to investigate the robustness of the  phenomenological models including the saturation phenomenon which have their parameters fixed by DESY-HERA data  when extrapolated to the very high energy regime reached at the LHC. 

Our goal in this work is to analyze the theoretical uncertainties on the predictions for  the photoproduction of light vector mesons in coherent $pp$, PbPb and $p$Pb  collisions at the LHC using the color dipole approach. Predictions for the rapidity distribution for $\rho^0$ and $\phi$ photoproduction will be furnished and an analysis on the uncertainties associated to the choice of vector meson wavefunctionand the phenomenological models for the dipole cross section is performed. Moreover, we will present our predictions for the rapidity dependence of $\rho$ cross sections at LHC energy of 2.76 TeV in coherent PbPb collisions, which is currently under analysis by the ALICE collaboration \cite{ALICErho}. 

\section{Theoretical framework}
\begin{figure*}[t]
\includegraphics[scale=0.5]{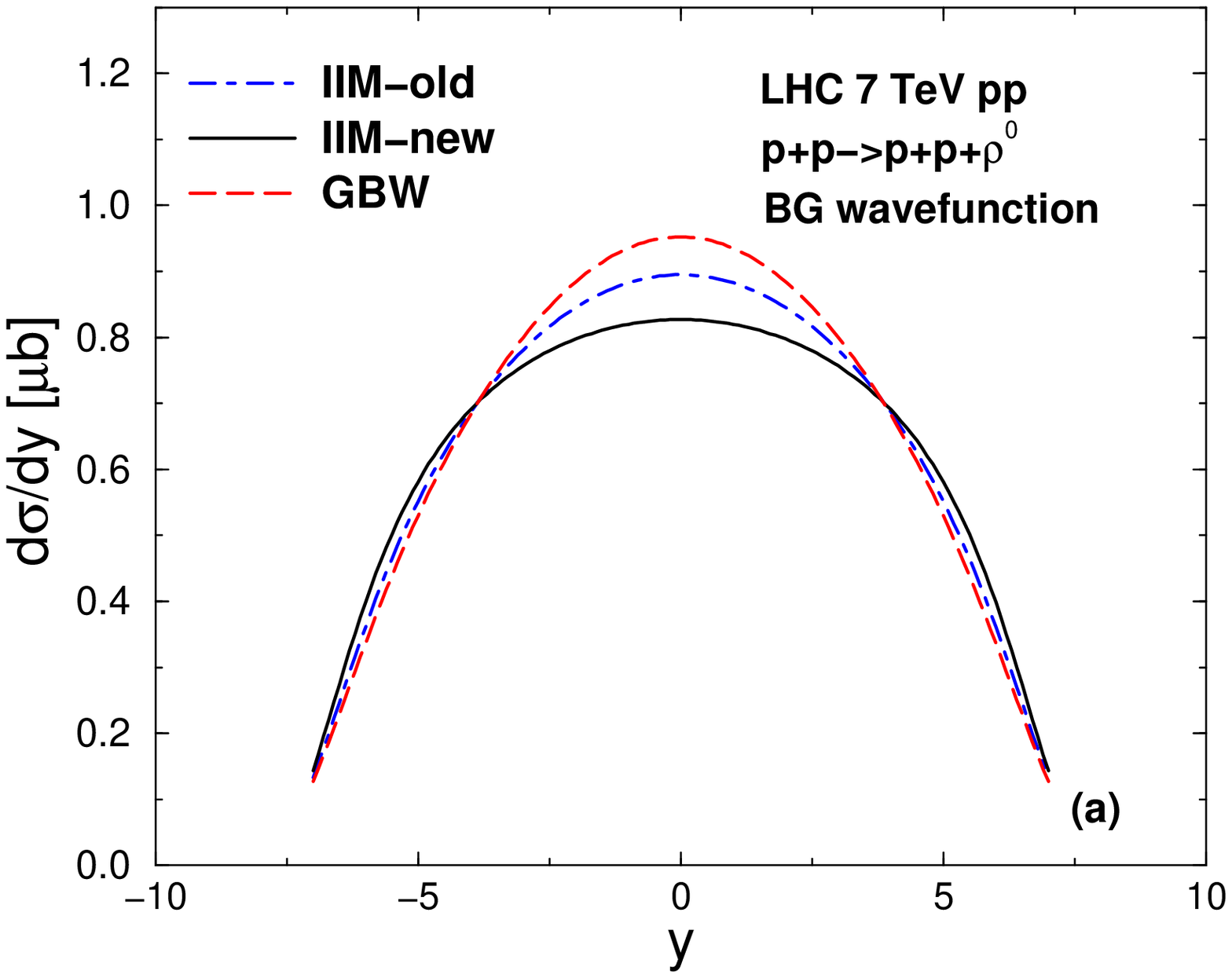} 
\includegraphics[scale=0.5]{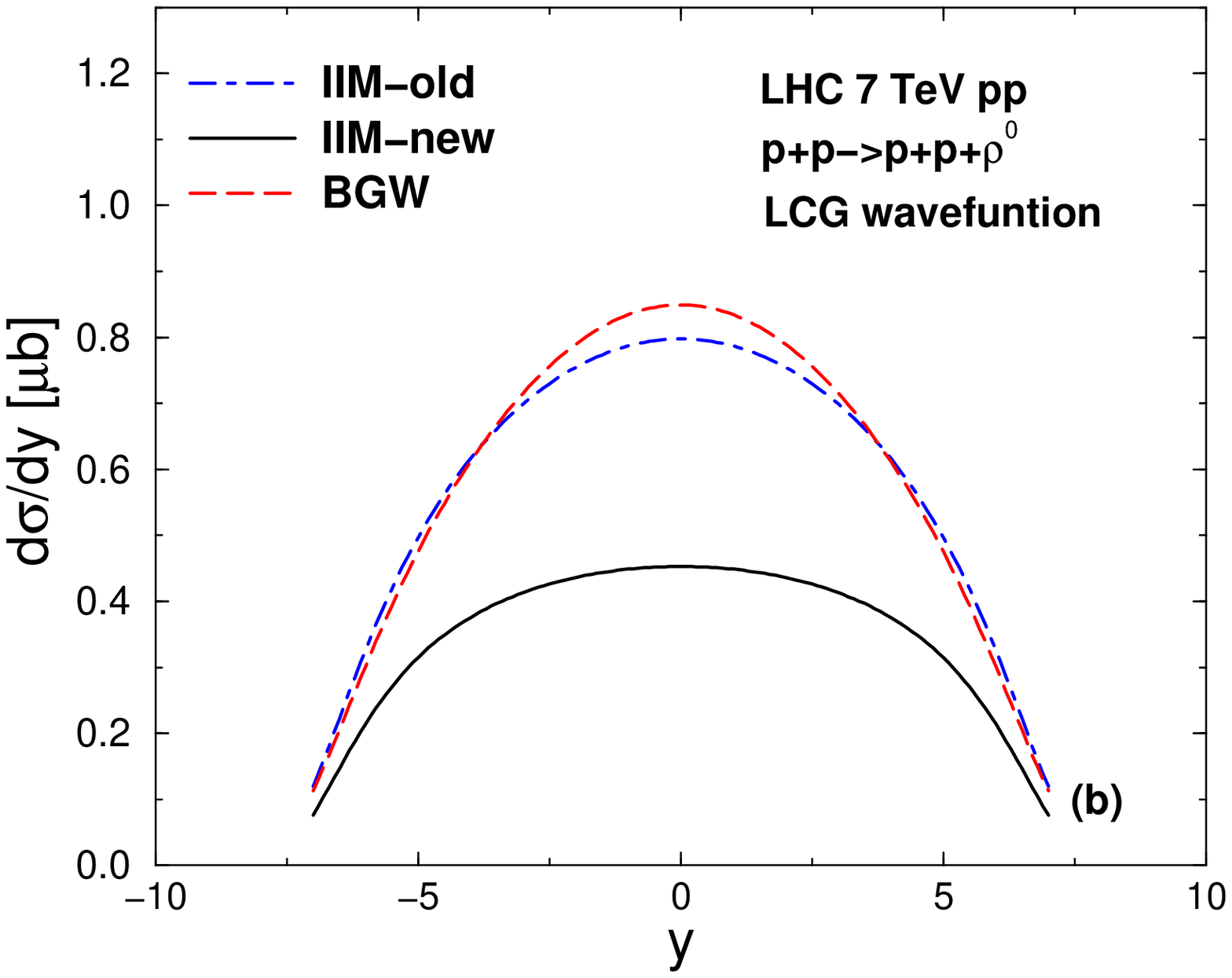}
 \caption{(Color online) Predictions for the rapidity distribution of $\rho^0$ photoproduction in $pp$ collisions at LHC ($\sqrt{s} = 7$ TeV) for the case of (a) Boosted Gaussian (BG) and (b) Light-Cone Gaussian (LCG) wavefunctions and several models for the dipole cross section (see text).}
\label{fig:1}
\end{figure*}

The exclusive meson photoproduction in hadron-hadron collisions can be factorized in terms of the equivalent flux of photons of the hadron projectile and photon-target production cross section \cite{upcs}. The photon energy spectrum, $dN_{\gamma}^p/d\omega$, which depends on the photon energy $\omega$,  is well known \cite{upcs}. The rapidity distribution $y$ for vector meson photoproduction in  $pp$ collisions can be  written down as,
\begin{eqnarray}
\frac{d\sigma}{dy}(pp \rightarrow   p\otimes V \otimes p) & = & S_{\text{gap}}^2 \left[\omega \frac{dN_{\gamma}^p}{d\omega }\sigma (\gamma p \rightarrow V p) \right. \nonumber \\
& + & \left.  \left(y\rightarrow -y \right) \right]. 
\label{dsigdy}
\end{eqnarray}
 The produced state with mass $m_V$ has rapidity $y\simeq \ln (2\omega/m_V)$ and the square of the $\gamma p$ center-of-mass energy is given by $W_{\gamma p}^2\simeq 2\omega\sqrt{s}$. The absorptive corrections due to spectator interactions between the two hadrons are represented by the factor $S_{\text{gap}}$. For simplicity here, we did not consider absorption corrections. The photon-Pomeron interaction will be described within the light-cone dipole frame, where the probing
projectile fluctuates into a
quark-antiquark pair with transverse separation
$r$ (and momentum fraction $z$) long after the interaction, which then
scatters off the hadron. The cross section for exclusive photoproduction of vector meson  off a nucleon target is given by \cite{Nemchik:1996pp} ,
\begin{eqnarray}
\label{sigmatot}
\sigma (\gamma p\rightarrow V p) & = &  \frac{1}{16\pi B_V} \left|\int dz\, d^2r \,\Phi^{\gamma^*V}_T\sigma_{dip}  \right|^2,\\
\Phi^{\gamma^*V}_T & = & \sum_{h, \bar{h}}\Psi^\gamma_{h, \bar{h}}(z,r,m_q)\Psi^{V*}_{h, \bar{h}}(z,r,m_q),
\end{eqnarray}
where $\Psi^{\gamma}(z,r,m_q)$ and $\Psi^{V}(z,r,m_q)$ are the light-cone wavefunction of the photon  and of the  vector meson, respectively.  The Bjorken variable is denoted by $x=M_V^2/(W_{\gamma p}^2-m_p^2)$, the dipole cross section by  $\sigma_{dip}(x,r)$ and the  diffractive slope parameter by $B_V$.  Here, we consider the energy dependence of the slope using the Regge motivated expression, $B_V(W_{\gamma p}) = B_0+4\alpha^{\prime}\log(W_{\gamma p}/W_0)$. We have considered $B_0=11$ GeV$^{-2}$, $W_0=95$ GeV and $\alpha^{\prime}=0.25$ GeV$^{-2}$ \cite{rhophotH1}. Similarly, the rapidity distribution $y$ in nucleus-nucleus collisions has the same factorized form,
\begin{eqnarray}
\frac{d\sigma}{dy} (A A \rightarrow   A\otimes V \otimes Y) & = &  \left[ \omega \frac{dN_{\gamma}^A}{d\omega }\,\sigma(\gamma A \rightarrow V +Y ) \right. \nonumber \\
& + & \left. \left(y\rightarrow -y \right) \right],
\label{dsigdyA}
\end{eqnarray}
where the photon flux in nucleus is denoted by $dN_{\gamma}^A/d\omega$ and $Y=A$ (coherent case) or $Y=A^*$ (incoherent case). The exclusive photoproduction off nuclei for coherent and incoherent processes can be simply computed in high energies where the large coherence length $l_c\gg R_A$ is fairly valid. The expressions for both cases are given by \cite{Boris},
\begin{eqnarray}
\sigma (\gamma A \rightarrow VA ) & = & \int d^2b\, \left|\langle \Psi^V|1-\exp\left[-\frac{1}{2}\sigma_{dip} T_A \right]|\Psi^{\gamma}\rangle \right|^2, \label{eq:coher} \nonumber \\
\sigma (\gamma A \rightarrow VA^* )  & = & \frac{1}{16\pi\,B_V}\int d^2b\,T_A \left|\langle \Psi^V|\sigma_{dip}(x,r) \right. \nonumber \\
 &\times &  \left. \exp\left[-\frac{1}{2}\sigma_{dip}T_A(b)  \right]|\Psi^{\gamma}\rangle\right|^2, \nonumber
\label{eq:incoh}
\end{eqnarray} 
where $T_A(b)= \int dz\rho_A(b,z)$  is the nuclear thickness function. The notation $\langle \Psi^V| (\ldots)|\Psi^{\gamma}\rangle$ represents the overlap over the wavefunctions. The rapidity distribution for the case of coherent $pA$ collisions can be also obtained. Disregarding the contribution from photonuclear interaction, the simplified expression is given by,
\begin{eqnarray}
\frac{d\sigma}{dy} (p A \rightarrow   p\otimes V \otimes A) & = &   \omega \frac{dN_{\gamma}^A}{d\omega }\,\sigma(\gamma A \rightarrow VA) 
\label{dsigdypA}
\end{eqnarray}

\begin{figure*}[t]
\includegraphics[scale=0.5]{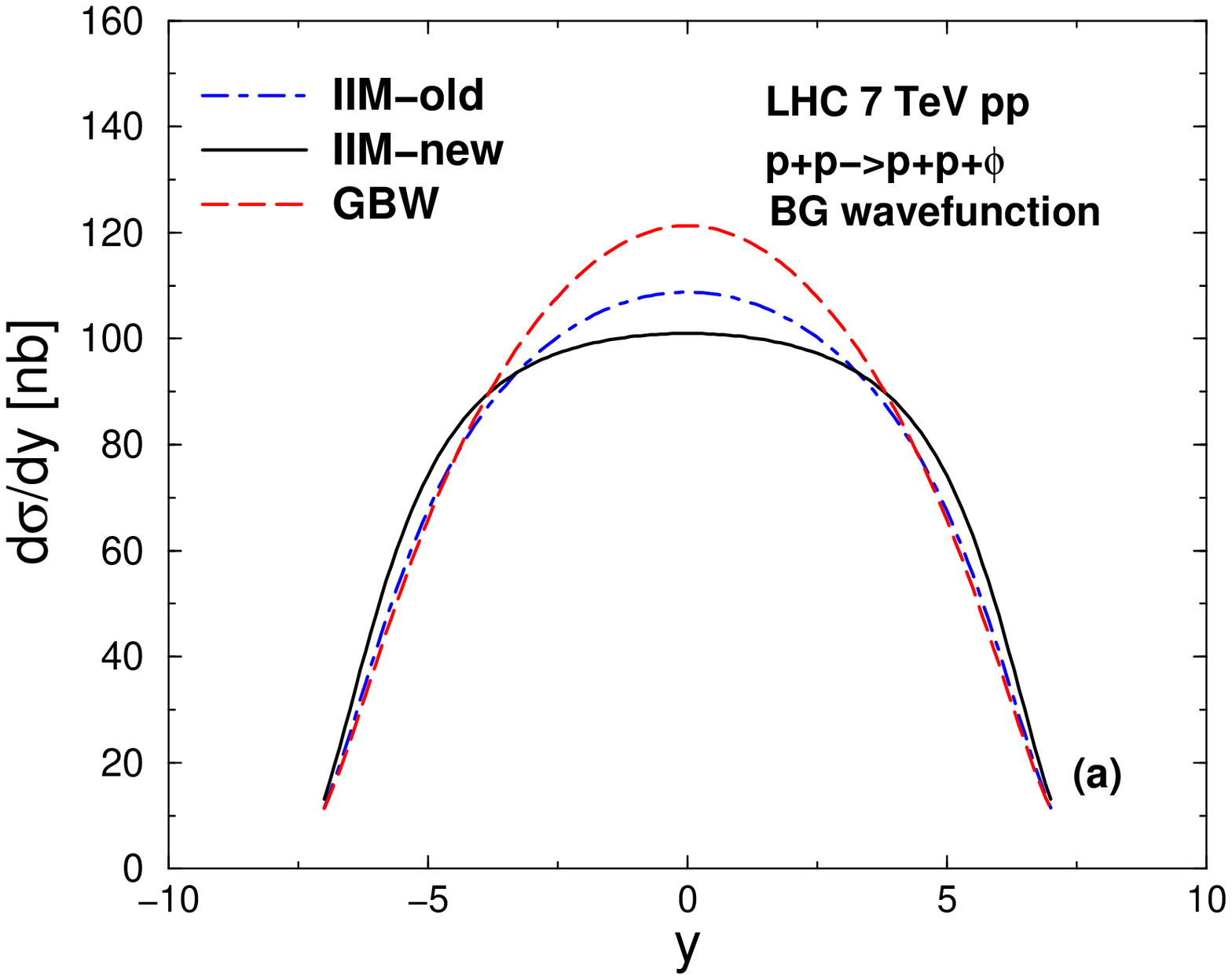} 
\includegraphics[scale=0.5]{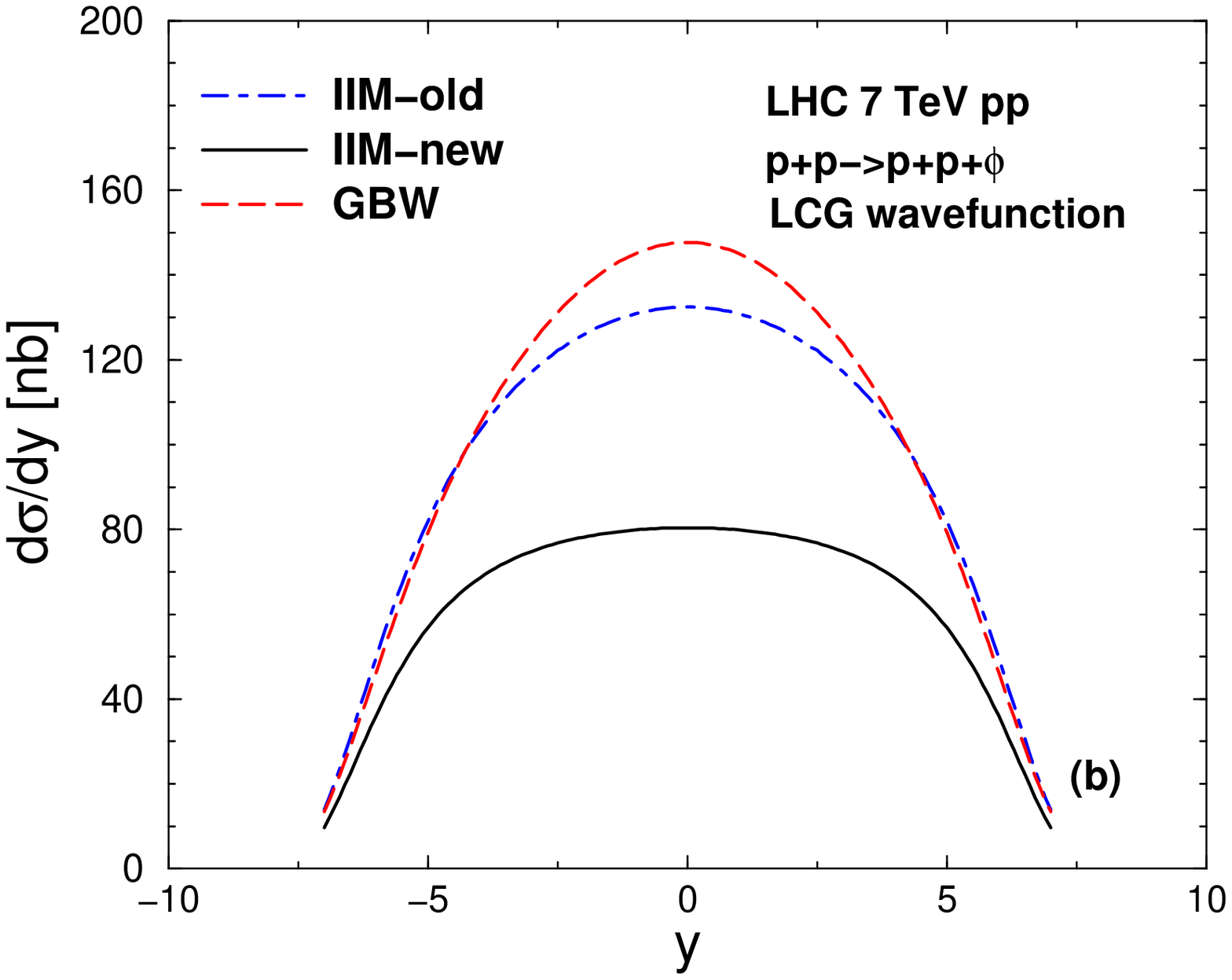}
 \caption{(Color online) Predictions for the rapidity distribution of $\phi$ photoproduction in $pp$ collisions at LHC ($\sqrt{s} = 7$ TeV) for the case of (a) Boosted Gaussian (BG) and (b) Light Cone Gaussian (LCG) wavefunctions and several models for the dipole cross section (see text).}
\label{fig:2}
\end{figure*}

In the numerical evaluations in next section, we have considered the Boosted Gaussian \cite{wfbg} (BG) and the Light-Cone Gaussian \cite{wflcg} (LCG) wavefunctions and the  phenomenological saturation models proposed in Ref. \cite{IIM} (IIM model) and Ref. \cite{GBW} (GBW model)  which encode the main properties of the saturation approaches.  The expressions for
the overlap functions we have used appropriately summed over the helicity and flavor indices are given by:
\begin{eqnarray*}
\Phi^{\gamma^*V}_T(z,r,m_q)  & = &  \hat{e}_f \frac{\sqrt{4\pi\,\alpha_e}}{(2\pi)^2} N_c\left\{
             m_f^2 K_0(r\epsilon_f)\phi_T(r,z) \right. \nonumber \\
           & - & \left. [z^2+(1-z)^2]\,\epsilon_f K_1(r\epsilon_f) \partial_r\phi_T(r,z)
            \right\},
\end{eqnarray*}
where the constant $\hat{e}_f$ stands for an effective charge. It is
given in Table \ref{tab:wfparams} along with the quark and meson
masses used. Here,  $m_f$ denotes the mass of the quark with flavor $f$ 
and with $\epsilon_f^2 = m_f^2$.  For the BG wavefunctions \cite{wfbg}, the function $\phi_T$ is given by,
\begin{eqnarray}
\phi_{T} = N_{T}\,4\sqrt{2\pi R^2}
\,\exp\left[-\frac{m_f^2R^2}{8z(1-z)}+\frac{m_f^2R^2}{2}-\frac{2z(1-z)r^2}{R^2}
\right]. \nonumber \\
\end{eqnarray}
The parameters $R$ and $N_{T}$ are constrained by unitarity of the
wavefuntion as well as by the electronic decay widths. They are given
in Table \ref{tab:wfparams} \cite{FSlight}. On the other hand, for LCG wavefunction
\cite{wflcg} one has the following expression:
\begin{eqnarray}
\phi_T  =  N_T\,z(1-z)\,\exp\left[-r^2/(2R_T^2)\right],
\end{eqnarray}
with the parameters also given in Table \ref{tab:wfparams} \cite{MPS}. The parameters for the meson wavefunctions shown in Table correspond to a fixed quark mass of $m_q=0.14$ GeV. We have verified that there is some sensitivity when considering a smaller quark mass, which leads to a change in the overall normalization $N_T$ (it increases as $m_q$ diminishes). 

The two wavefunctions considered above are samples of the available phenomenological models. For instance, for $\rho$ production Forshaw and Sandapen \cite{FS1,FS2,FS3} have used the state of art to constraint the parameters of meson wavefunctionfrom recent data on rho electroproduction. In Ref. \cite{FS1}  the $\rho$ wavefunctionwas extracted from HERA data and it was found that they prefer a transverse wavefunction with enhanced end-point contributions. In Ref. \cite{FS2} the leading twist-2 and sub-leading twist-3 Distribution Amplitudes of the rho meson have been extracted and in Ref. \cite{FS3} they provided an AdS/CFT holographic wave-funtion for the rho meson and compare it to the available data on photon-proton process. 

\begin{table}[h]
\begin{center}
\begin{tabular}{|c|c|c|c|c|c|c|c|}
\hline
             & \multicolumn{3}{|c|}{common parameters}
             & \multicolumn{2}{|c|}{BG parameters}
             & \multicolumn{2}{|c|}{LCG parameters}\\
\hline
$V$  & $M_V$ & $m_f$ & $\hat{e}_f$ &  $R^2$  & $N_T$ 
             & $R_T^2$  & $N_T$ \\
\hline
$\rho$  &   0.776 &  0.14 (0.01) & 1/$\sqrt{2}$ & 12.3 & 0.0259 & 21.0 & 4.47 \\
$\phi$  &   1.019 &  0.14 (0.01) & 1/3 & 10.0 & 0.0251 &  16.0 &   4.75 \\
\hline
\end{tabular}
\end{center}
\caption{Parameters for the vector-meson light-cone wavefunctions \cite{FSlight,MPS} in units of GeV.}
\label{tab:wfparams}
\end{table}

For the phenomenological models for the dipole-proton cross section, we have considered two set of parameters for the IIM parameterization \cite{IIM} (including charm quark in fits). In this case, the dipole cross section  is parameterized as follows,
\begin{eqnarray}
\sigma_{dip}\,(x,r) =\sigma_0\,\left\{ \begin{array}{ll} 
0.7 \left(\frac{\bar{\tau}^2}{4}\right)^{\gamma_{\mathrm{eff}}\,(x,\,r)}\,, & \mbox{for $\bar{\tau} \le 2$}\,,  \nonumber \\
 1 - \exp \left[ -a\,\ln^2\,(b\,\bar{\tau}) \right]\,,  & \mbox{for $\bar{\tau}  > 2$}\,, 
\end{array} \right.
\label{CGCfit}
\end{eqnarray}
where $\bar{\tau}=r Q_{\mathrm{sat}}(x)$ and the expression for $\bar{\tau} > 2$  (saturation region)   has the correct functional
form, as obtained  from the theory of the Color Glass Condensate (CGC) \cite{hdqcd}. For the color transparency region near saturation border ($\bar{\tau} \le 2$), the behavior is driven by the effective anomalous dimension $\gamma_{\mathrm{eff}}\, (x,\,r)= \gamma_{\mathrm{sat}} + \frac{\ln (2/\tilde{\tau})}{\kappa \,\lambda \,y}$ with $\kappa = 9.9$. The saturation scale is defined as $Q_{sat}^2(x)=\left(\frac{x_0}{x}\right)^{\lambda}$ and $\sigma_0=2\pi R_p^2$.

\begin{figure*}[t]
\includegraphics[scale=0.5]{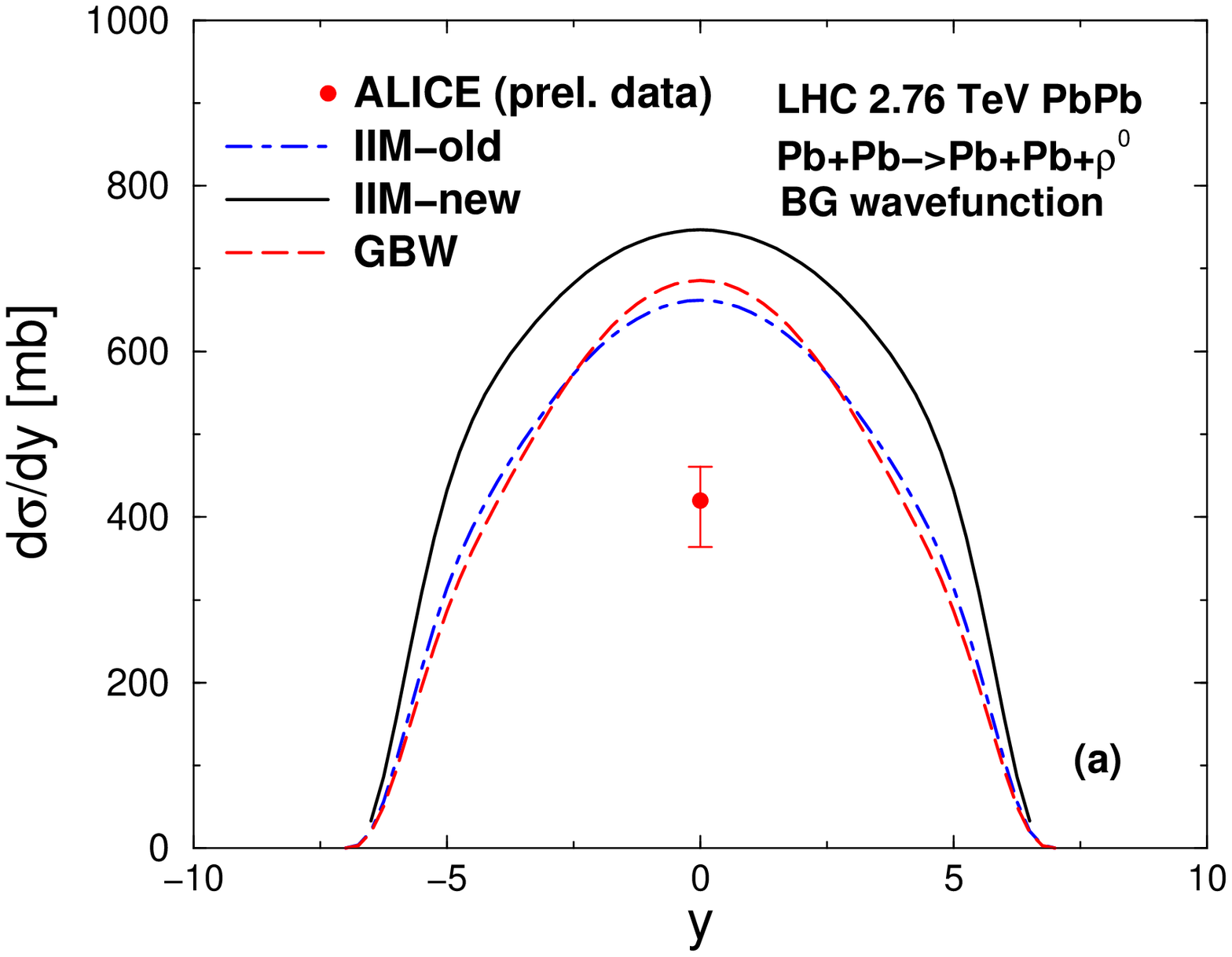} 
\includegraphics[scale=0.5]{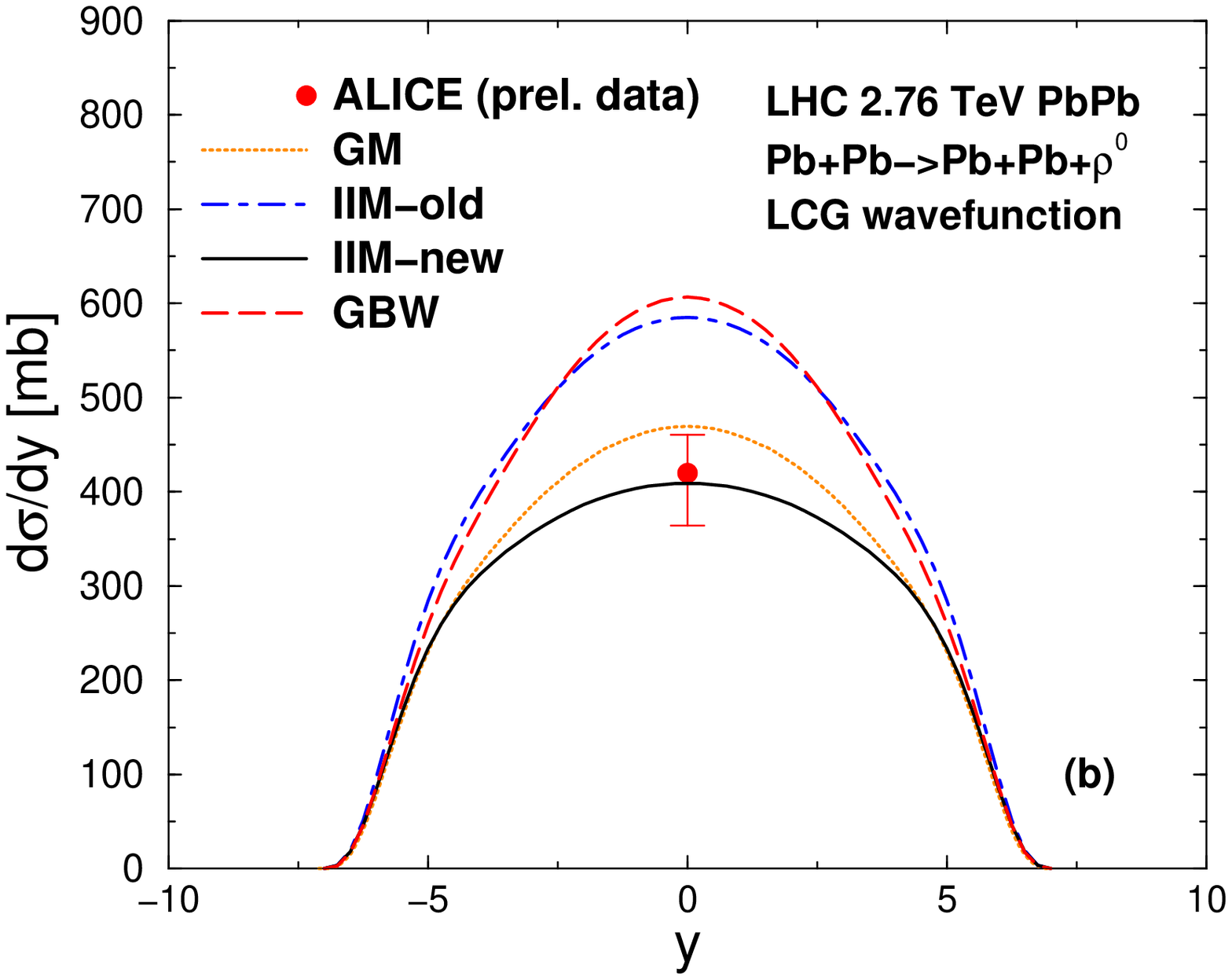}
 \caption{(Color online) Predictions for the rapidity distribution of $\rho^0$ photonuclear production in PbPb collisions at LHC ($\sqrt{s} = 2.76$ TeV) for the case of (a) Boosted Gaussian (BG) and (b) Light Cone Gaussian (LCG) wavefunctions and several models for the dipole cross section (see text). The preliminary ALICE data \cite{ALICErho} for central rapidity is also presented.}
\label{fig:3}
\end{figure*}

The first set (labeled by IIM-old \cite{Soyez}) considers the previous DESY-HERA data and the values for parameters are $\gamma_{\mathrm{sat}}=0.7376$, $\lambda = 0.2197$, $x_0=0.1632\times 10^{-4}$ and $R_p= 3.344$ GeV$^{-1}$ ($\sigma_0=27.33$ mb). For IIM-old, the light quark mass is fixed as $m_{u,d,s}=0.14$ GeV. The second set (labeled IIM-new \cite{Amir}) considered the extremely small error bars on the recent ZEUS and H1 combined results for inclusive DIS. In this case, the parameters are $\gamma_{\mathrm{sat}}=0.762$, $\lambda = 0.2319$, $x_0=0.6266\times 10^{-4}$ and $\sigma_0=21.85$ mb. For IIM-new, the light quark mass is fixed as $m_{u,d,s}=10^{-2}-10^{-4}$ GeV (we take $m_q=0.01$ in the numerical calculations). At the same $x$-value, the saturation scale is higher for IIM-new as $x_0$ and $\lambda$ are both bigger as for IIM-old case. On the other hand, the asymptotic value of dipole cross section $\sigma_0$ is smaller for IIM-new. We also see that it is obtained a smaller $\Phi^{\gamma^*V}_T(z,r,m_q)$, which is directly dependent on $m_q$, when compared to value of mass $m_q=0.14$ GeV.

In order to compare the dependence on distinct models, we also consider the simple GBW parametrization \cite{GBW}, where the dipole cross section is given by,
\begin{equation}
\sigma_{dip}(x,r)=\sigma_{0}\left[ 1-\exp\left(-\frac{r^{2}Q_{\mathrm{sat}}^{2}}{4}\right)^{\gamma_{\mathrm{eff}}}\,\right],\label{gbw}
\end{equation}
where the the effective anomalous dimension is taken as  $\gamma_{\mathrm{eff}} = 1$.

As a final note on the details of the present calculation, we discuss the threshold correction, the real part of amplitude and skewdness effects. In all numerical calculations, we multiply the dipole cross sections above by a threshold correction factor $(1-x)^n$, where $n = 5$ for light mesons and $n = 7$ for the heavy ones (the value for $n$ is estimated using quark counting rules). The cross section in Eq. (\ref{sigmatot}) has been computed including the real part of amplitude contribution and skwedness correction in the following way, 
\begin{equation}
 \hat{\sigma}_{\gamma p\rightarrow Vp}=R_g^{2} \,{\sigma}(\gamma p\rightarrow Vp)(1+\beta^2) ,
  \label{vm}
\end{equation}
with
\begin{eqnarray}
 \label{eq:beta} 
  \beta & = &  \tan\left(\frac{\pi\varepsilon }{2}\right), \hspace{0.6cm}R_g(\varepsilon) = \frac{2^{2\varepsilon+3}}{\sqrt{\pi}}\frac{\Gamma(\varepsilon+5/2)}{\Gamma(\varepsilon+4)}, \nonumber\\
\varepsilon &\equiv& \frac{\partial\ln\left(\mathcal{A}^{\gamma p\rightarrow Vp}\right)}{\partial\ln(1/x)}, \
\end{eqnarray}
where the factor $(1+\beta^2)$ takes into account the missing real part of amplitude, with $\beta$ being the ratio of real to imaginary parts of the scattering amplitude.  The factor $R_g$ incorporates the skewness effect, coming from the fact that the gluons attached to the $q\bar{q}$ can carry different light-front fractions $x,x^{\prime}$ of the proton.
The skewedness factor given in Eq. (\ref{eq:beta})  was obtained at NLO level, in the limit that $x^{\prime}\ll x\ll 1$ and at small $t$ assuming that the diagonal gluon density of target has a power-law form  \cite{skew}.

\section{Results and discussions}

\begin{figure}[t]
\includegraphics[scale=0.5]{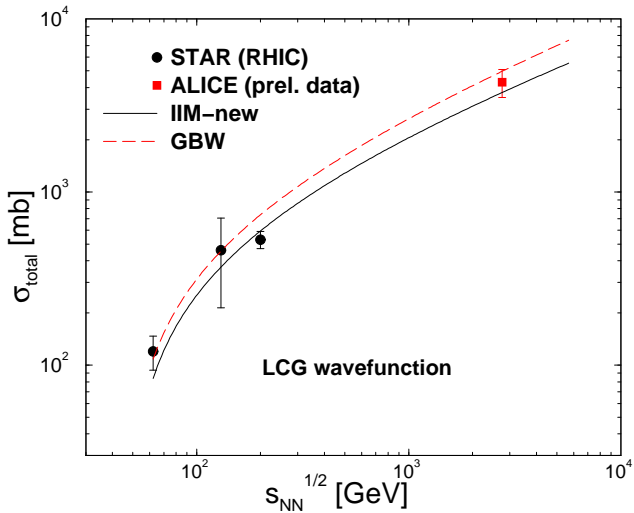}
 \caption{(Color online) The total cross section for $\rho^0$ coherent production in nucleus-nucleus collisions as a function of energy $\sqrt{s_{NN}}$. The curves are for IIM-model (solid line) and GBW (dashed line) models for dipole cross section and both considering LCG wavefunction. The measurements of STAR (RHIC) \cite{RHIC} at low energy and the preliminary ALICE data \cite{ALICErho} are also presented.}
\label{fig:4}
\end{figure}

Let us start by calculating the rapidity distribution for $\rho$ and $\phi$ production in proton-proton collisions at the energy of 7 TeV. In Fig. \ref{fig:1}-(a)  is presented the results for $\rho^0$ taking into account the Boosted Gaussian (BG) wavefunction and some samples of phenomenological models for the dipole cross section. The dot-dashed curve stands for the IIM dipole cross section using previous values of its fitted parameters (IIM-old). The solid line represents the result using the new fitted parameters for IIM model (IIM-new), whereas the dashed curve stands for the celebrated GBW parametrisation. The behavior at large rapidities is similar for the distinct models. However, at mid-rapidities there is an evident model dependence. It is found $d\sigma /dy(y=0)= 0.9,\,0.83,\,0.95$ $\mu$b for IIM-old, IIM-new and GBW, respectively. The deviation is order 14\% in that case. In Fig. \ref{fig:1}-(b), the results are now presented for the Light Cone Gaussian (LCG) wavefunction and the notation is the same as for the previous plot. In the mid-rapidity region one gets  $d\sigma /dy(y=0)= 0.80,\,0.45,\,0.85$ $\mu$b for IIM-old, IIM-new and GBW, respectively. The predictions using the LCG wavefunction are smaller than the BG wavefunction  case. In addition, there is an intense suppression when IIM-new model is considered (a reduction by a factor 1.8). Concerning the $\phi$ meson production, in Fig. \ref{fig:2}-(a) one presents the results using the BG wavefunction and in Fig.  \ref{fig:2}-(b) the corresponding values for LCG wavefunction. The notation is that same as the plots for the $\rho$ case. At mid-rapidity it is found $d\sigma /dy(y=0)=$ 108.8  nb (IIM-old), 101  nb (IIM-new) and  121.3 nb (GBW) using the BG wavefunction and $d\sigma /dy(y=0)=$ 132.4  nb (IIM-old), 80.4 nb (IIM-new) and  147.7 nb (GBW)  considering the LCG wavefunction. In contrast with the $\rho$ case, the mid-rapidity value of cross sections are higher using LCG instead of BG wavefunction by a factor 20\% at least for IIM-old and GBW models. This can be due the richer structure of the BG wavefunction in comparison to the LCG wavefunction. Once again, a reduction is observed using the IIM-new model. As predictions for the 14 TeV run, it is found $d\sigma_{\rho} /dy(y=0)=0.71\pm 0.21$ $\mu$b (IIM-new) and  $d\sigma_{\rho} /dy(y=0)=1.04\pm 0.06$ $\mu$b (GBW). The errors take into account the dependence on the wavefunction. For the $\phi$ case we get $d\sigma_{\phi} /dy(y=0)=99\pm 11$ nb (IIM-new) and  $d\sigma_{\phi} /dy(y=0)=153\pm 16$ nb (GBW). The large theoretical uncertainty presented here it was also found when considering the pQCD $k_{\perp}$-factorization approach \cite{Szczurek1,Szczurek2}, where the authors also considered the absorption effects. Our predictions are somewhat consistent with those in Refs. \cite{Szczurek1,Szczurek2} at mid-rapidity for 14 TeV.

Now, we investigate the photonuclear production of $\rho$ and $\phi$ mesons in nucleus-nucleus collisions at the LHC. We will consider PbPb  collisions at the energy of 2.76 TeV. In Fig. \ref{fig:3}-(a) we present the results for the rapidity distributions for the coherent $\rho$ production, $\mathrm{Pb}+\mathrm{Pb}\rightarrow \mathrm{Pb}+\rho^0+\mathrm{Pb}$, considering the BG wavefunction (without nuclear break up). The preliminary ALICE data \cite{ALICErho} for coherent $\rho$ production, $\frac{d\sigma}{dy}(y=0)= 420 \pm 10\,(\mathrm{stat.}) \,^{+30}_{-55}\,(\mathrm{syst.})$ mb,  is also presented and the  notation for the curves is the same as for the proton-proton case. It was found  $d\sigma /dy(y=0)=$ 661.5 mb (IIM-old),  747 mb (IIM-new) and 685.6 mb (GBW), respectively. In any case, the predictions are in average 50\% larger than the experimental result. The theoretical uncertainty associated to the model for the dipole cross section remains as in the proton-proton case. It is a distinction compared to the proton-proton case for the IIM-new for BG wave-function, where the prediction is larger than IIM-old and GBW. A careful analysis on the quark mass dependence for the BG wavefunction would be in order. In Fig. \ref{fig:3}-(b) one presents the results considering the LCG wavefuntion, including the previous BG prediction \cite{BG} (dotted line) that also considered the color dipole approach. This time it is obtained the values  $d\sigma /dy(y=0)=$ 469.5 mb (GM), 585.1 mb  (IIM-old),  409 mb (IIM-new) and 603.3 mb (GBW), respectively. The results are smaller that for the BG wavefunction and the IIM-new result is consistent with data within the error bars.  In Fig. \ref{fig:4} we present the integrated cross section (all rapidities) as a function of $NN$-energy, including the lower energy ($\sqrt{s_{NN}}=$ 62.4, 130 and 200 GeV) results from STAR Collaboration at RHIC \cite{RHIC}. The preliminary data from ALICE, $\sigma_{total}^{coh} = 4.3 \pm 0.1\,(\mathrm{stat.}) \,^{+0.6}_{-0.5}\,(\mathrm{syst.})$ b is also presented. It is shown the predictions using the IIM-new (solid line) and GBW (dashed line) models for the dipole cross section and the LGC wavefuntion. When comparing the models investigated here to those data we have presented the coherent cross section only (we did not include contributions with nuclear break up \cite{Baltz})  and a interpolation from RHIC to LHC is performed.  It is verified that the color dipole approach gives a reasonable description of energy dependence from low to high energies. As a prediction for the future 5.5 TeV run, we obtain the following the cross sections $\sigma_{0n0n}^{coh}(\mathrm{all}\,y) = 5.30\, (7.21)$ b using the LCG wavefunction and models IIM-new (GBW) for the dipole cross section. Our prediction are smaller (for this particular choice of wavefunction) than those presented in Ref. \cite{RSZ} (RSZ) where it is found $\sigma_{0n0n}^{\mathrm{RSZ}}(\mathrm{all}\,y) = 8.309$ b at 5.5 TeV.

\begin{figure}[t]
\includegraphics[scale=0.5]{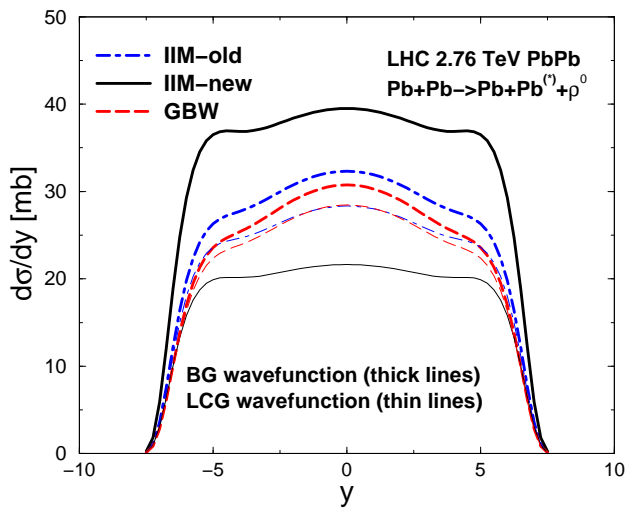}
 \caption{(Color online) The rapidity distribution  for $\rho^0$ incoherent production in PbPb collisions at 2.76 TeV. The thick lines represent the predictions using BG wavefunction and the thin lines represent the predictions for LCG wavefunction. }
\label{fig:5}
\end{figure}

\begin{figure}[t]
\includegraphics[scale=0.5]{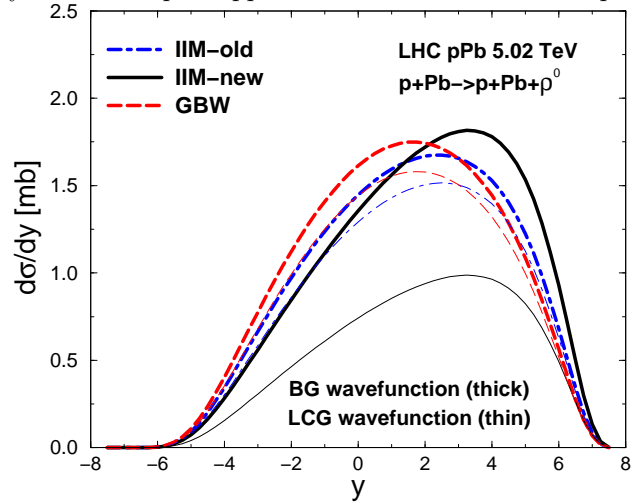}
 \caption{(Color online) The rapidity distribution  for $\rho^0$ coherent production in pPb collisions at 5.02 TeV. The thick lines represent the predictions using BG wavefunction and the thin lines represent the predictions for LCG wavefunction. }
\label{fig:6}
\end{figure}

In Fig. \ref{fig:5} it is presented the results for the incoherent $\rho^0$ photonuclear production, $\mathrm{Pb}+\mathrm{Pb}\rightarrow \mathrm{Pb}+\rho^0+\mathrm{Pb}^{(*)}$, using same notation as previous plots. In a similar way as for the coherent case, the predictions using BG wavefunction (thick lines) are larger that the LCG option (thin lines)  and can introduce an uncertainty by a factor two (see IIM-new case) at mid-rapidities. The main prediction is that the incoherent $\rho$ production is of order $30\pm 10$ mb at $y=0$ for energy of 2.76 TeV. Finally, the rapidity distribution for the $pA$ interaction is shown in Fig. \ref{fig:6}, where the results for BG (thick lines) and LCG (thin lines)  wavefunctions are presented in a single plot. We can see a strong dependence on the choice of meson wavefunction and on the  dipole cross section, where the smaller cross section is provided by the IIM-new parameterization and LCG wavefunction. The pattern presented for the $\rho$ production is similar to the results for quarkonium production recently investigated in Refs. \cite{GM,GMN} using also the color dipole approach.

As a summary, in this work we performed calculations considering the color dipole approach leading to predictions for the light vector meson production as $\rho$ and $\phi$  in coherent and incoherent interactions at the LHC energies for $pp$, $pPb$ and PbPb collisions. We show that the theoretical uncertainty is considerably large and the main sources are the models for the meson wavefunction and the phenomenological models for the dipole cross section. The BG wavefunction leads to larger cross sections at mid-rapidity compared tho the LCG wavefunction and the dependence of the overall normalization with the color dipole model is important. In particular, the recent ALICE preliminary data seems to favor the LCG wavefunction and more recent parameterizations for the dipole cross section.   Our results demonstrate that the production rates in RHIC and LHC are fairly described by the color dipole approach. This corroborates the previous studies on heavy meson production as $J/\psi$ and $\psi (2S)$ \cite{GDGM} (see also Fig. 7 from Ref. \cite{LHCb2V}) using the same framework presented here.

\begin{acknowledgments}
This work was  partially financed by the Brazilian funding agency CNPq and by the French-Brazilian scientific cooperation project CAPES-COFECUB 744/12. MVTM thanks Amir Rezaeian for useful discussions.
\end{acknowledgments}

\end{document}